\documentstyle[mnextra]{mn}
\seceqnum           
\input epsf.tex
\def\etal{{\it et al.} }
\def\kms  {{\rm km \, s^{-1}}}
\def\Mpc  {{\rm h^{-1}  Mpc}}
\def\kpc  {{\rm h^{-1}  kpc}}
\def\Msol {{\rm h^{-1}  M_\odot}}
\def\d    {{ \rm d}}
\def\eg   {{e.g.\ }}
\def\dcz  {\delta_{\rm c}(z)}
\def\lsim{\mathrel{\hbox{\rlap{\hbox{\lower4pt\hbox{$\sim$}}}\hbox{$<$}}}}
\def\gsim{\mathrel{\hbox{\rlap{\hbox{\lower4pt\hbox{$\sim$}}}\hbox{$>$}}}}

\begin{document}
\title[Cluster Evolution]
{Using the Evolution of Clusters to Constrain $\Omega$}
\author[V.R.Eke, S.Cole and C.S.Frenk ]{
Vincent R. Eke$^{1,2}$, Shaun Cole$^{1,3}$ and Carlos S. Frenk$^{1,4}$ \\
$^{1}$Department of Physics, University of Durham, Science
Laboratories, South Rd, Durham DH1 3LE\\
$^{2}$V.R.Eke@durham.ac.uk\\
$^{3}$Shaun.Cole@durham.ac.uk\\
$^{4}$C.S.Frenk@durham.ac.uk}

\maketitle

\begin{abstract}

The population of rich galaxy clusters evolves much more rapidly in a
universe with critical density than in a universe with low density. Thus,
counts of clusters at intermediate redshift offer the possibility of
determining the cosmological density parameter, $\Omega_0$, with a minimum
of assumptions. We quantify this evolution using the Press-Schechter
formalism which we extend to flat cosmological models with a cosmological
constant, $\Lambda_0=1-\Omega_0$. Using new large N-body simulations, we
verify that this formalism accurately predicts the abundance of rich
clusters as a function of redshift in various cosmologies. We normalise the
models by comparing them to the local abundance of clusters as a function of
their X-ray temperature which we rederive from data compiled by Henry \&
Arnaud. The resulting values of the {\it rms} density fluctuation in
spheres of radius $8\Mpc$ are $\sigma_8 = (0.50 \pm 0.04)
\Omega_0^{-0.47+0.10\Omega_0}$ if $\Lambda_0=0$ and $\sigma_8 =
(0.50 \pm 0.04) \Omega_0^{-0.53+0.13\Omega_0} $ if $\Lambda_0=1-\Omega_0$.
These values depend only weakly, and almost not at all if $\Omega_0=1$, on the
shape of the power spectrum. We then examine how the distributions of mass,
X-ray temperature and Sunyaev-Zel'dovich decrement evolve as a function of
$\Omega_0$. We present the expected distributions at $z=0.33$ and $z=0.5$
and the predicted number counts of the largest clusters, both in space and
in projection on the sky. 
We find that even at $z=0.33$, these distributions 
depend very strongly on $\Omega_0$ and only weakly on 
$\Lambda_0$. For example, at this redshift, we expect 20 times as many
clusters per comoving volume with $M>3.5 \times 10^{14}\Msol$ and 5 times
as many clusters with $kT>5$ keV if $\Omega_0=0.3$ than if
$\Omega_0=1$. The splitting in the integrated counts is enhanced by the
larger volume element in low $\Omega_0$ models.
There is therefore a real prospect of estimating $\Omega_0$ from forthcoming 
surveys of intermediate redshift clusters that will determine their masses,
X-ray temperatures or Sunyaev-Zel'dovich decrements. 

\end{abstract}
\begin{keywords}
galaxies: clusters -- cosmology: theory .
\end{keywords}

\section{Introduction}

Galaxy clusters are exceptionally useful tools for estimating fundamental
cosmological parameters. Their utility stems largely from their relative
dynamical youth. In hierarchical clustering theories, clusters form by the
gravitational amplification of primordial density fluctuations, usually
assumed to have an initially Gaussian distribution of amplitudes.  Rich
clusters correspond to the rarest collapsed objects in this distribution
and their abundance varies very rapidly with properties such as mass or
potential well depth.

The mass within the virial radius of a rich cluster ($\sim 5\times
10^{14}\Msol$) is very close to the mass enclosed within a sphere of radius
$8 \Mpc$ in the unperturbed universe (Evrard 1989, White, Efstathiou \&
Frenk 1993). Because of this, the present day abundance of rich clusters
directly reflects the amplitude of density fluctuations on a scale of $\sim
8\Mpc$ and can be used to measure this amplitude with a minimum of
assumptions.  In general, this measure depends only weakly 
on the amplitude of fluctuations on other scales. It does,
however, depend on the value of the cosmological density parameter,
$\Omega_0$. Thus, the observed local cluster abundance fixes the value of
$\sigma_8$, the {\it rms} density fluctuation in spheres of radius $8\Mpc$,
as a function of $\Omega_0$.

The temporal {\it evolution} of the cluster abundance is determined by the
rate at which density perturbations grow. This, in turn, depends
primarily on the value of $\Omega$ and, to a lesser extent, the value of
the cosmological constant, $\Lambda$\footnote{We express $\Lambda$ in units
of $3H_0^2$, where $H_0= 100 {\rm h} \kms {\rm Mpc}^{-1}$ is the present
value of the Hubble constant}, and the shape of the power spectrum of
density fluctuations. In a low density universe, fluctuations cease to grow
after a redshift $z \sim (\Omega_{0}^{-1}-1)^{-1}$ (\eg Peebles 1980), 
resulting in a cluster population that evolves very
slowly at low redshift. In an $\Omega_0=1$ universe, on the other hand,
density fluctuations continue to grow even at the present epoch and so the
cluster population is still evolving rapidly. Measurement of the cluster
abundance at moderate redshifts provides a powerful method for estimating 
$\Omega_0$.

In order to apply these tests, it is necessary to predict and measure the
abundance of clusters as a function of some property such as mass, X-ray
luminosity or the equilibrium temperature of the X-ray emitting
intracluster gas. Theoretical predictions and observational measurements
are subject to different uncertainties. The cluster mass within the virial
radius is the simplest quantity to predict theoretically, but one of the
hardest to measure reliably. Traditional virial analyses are prone to 
contamination by projection effects (Frenk \etal 1990; Van Haarlem, Frenk
\& White 1996; Mazure \etal 1995) and X-ray data do not generally
extend to the virial radius. A novel and highly promising technique to
measure cluster masses employs the shape distortions of background
galaxies produced by weak gravitational lensing in the cluster
potential (Kaiser \& Squires 1993, Fahlman \etal 1994,
Wilson, Cole \& Frenk 1996, Seitz \& Schneider 1995 and references therein.)
The limited field of view of the current generation of CCD cameras,
however, restricts such measurements to the inner parts of clusters.

The simplest quantity to measure empirically is the cluster X-ray
luminosity. However, since the bremsstrahlung emissivity per unit volume is
proportional to the square of the gas density, the total power radiated is
very sensitively dependent upon the distribution of gas in the cluster
core. This is difficult to model, particularly at high
redshift (Evrard 1990; Navarro, Frenk \& White 1995 and references
therein). Nevertheless, X-ray luminosity provides a convenient means to
select complete samples of galaxy clusters. In contrast, the temperature of
the intracluster gas can be predicted quite reliably using modern
hydrodynamic simulations. These show that as a cluster
collapses, the gas is shock heated to the virial temperature and rapidly
settles into hydrostatic equilibrium with an approximate isothermal
structure (Evrard 1990, Cen \& Ostriker 1994, Bryan \etal 1994,
Navarro \etal 1995). Average (luminosity-weighted) X-ray
temperatures have now been measured for fairly large samples of clusters, with
the Einstein Observatory \cite{ha}, and EXOSAT \cite{edget}, 
and the radial variation of
temperature is now beginning to be probed with ASCA \cite{asca}. Early 
results show that the gas is indeed approximately isothermal. Finally,
another observable that is also insensitive to the detailed distribution of
the gas within the cluster is the $\Delta T/T$ decrement in the cosmic
microwave background radiation produced by the Sunyaev-Zel'dovich (S-Z)
effect (Sunyaev \& Zeldovich 1972). The line-of-sight decrement from the 
cluster depends only
on the product of the column density of gas and the temperature and
is independent of the cluster redshift. Sunyaev-Zel'dovich decrements have
now been reliably measured for several clusters (eg Grainge \etal (1993),
Wilbanks \etal (1994) and Birkinshaw \& Hughes (1994)).

In this paper, we exploit those cluster properties which are best
determined theoretically and observationally in order to constrain the
values of cosmological parameters. We first use the distribution of X-ray
temperatures at the {\it present day} to estimate $\sigma_8$. We then
examine the {\it evolution} of cluster properties to constrain the value
of $\Omega_0$. Specifically, we present predictions for the mass function,
the temperature function and the distribution of S-Z decrements at $z\simeq
0.3-0.5$. This range of redshifts is a rich area of observational
work. Ongoing programmes include mass measurements from weak gravitational
lensing using HST and large format CCDs (Ellis, private communication;
Kaiser, private communication); estimates of the cluster temperature
function (Henry, private communication); and S-Z source counts and redshift
distributions from ground-based radio telescopes and, if approved, from
the COBRA-SAMBA satellite mission.

Related calculations have been carried out by Evrard (1989),
Frenk \etal (1990), Bond \& Myers (1992), Lilje (1992), Oukbir \&
Blanchard (1992), Bahcall \& Cen (1993), Hanami (1993), White \etal (1993), 
Bartlett \& Silk (1993,1994), 
Barbosa \etal (1995), Hattori \& Matsuzawa (1995), 
Viana \& Liddle (1995), amongst others. Our work differs from these earlier
studies in various respects. Firstly, we consider a wider range of 
cosmological models than most previous analyses, particularly models
with low values of $\Omega_0$, with and without a cosmological constant. 
Secondly, we use some of the largest N-body simulations ever carried out in
order to verify the accuracy of our analytical calculations. Finally, we 
place special emphasis on making detailed predictions in a manner that can
be compared to observational data with a minimum of assumptions.
Since high-redshift data are not yet sufficiently complete, we do not
attempt to estimate $\Omega_0$ in this paper, but present instead a number
of distributions that will enable this estimate to be made as the
data become available. 

This paper is structured as follows. Section~\ref{sec:model} reviews the
Press-Schechter model for the abundance of clusters, extended to open and
flat cosmological models ($\Omega_0<1$ with $\Lambda_0=0$ or $\Lambda_0= 
1-\Omega_0$).  The
predictions of this model are compared with a set of $192^3$ particle
N-body simulations in Section~\ref{sec:sims}.  In Section~\ref{sec:norm}
we review the existing cluster abundance data and normalise the analytical
models using the cumulative cluster temperature function estimated from
the data compiled by Henry \& Arnaud (1991). The evolution of the cluster
mass and temperature functions, the S-Z source counts and the dependence of
these quantities on $\Omega_0$ and $\Lambda_0$ are predicted in
Sections~\ref{sec:mevol}~-~\ref{sec:yevol}.  Finally, we discuss the
importance and implications of these results in Section~\ref{sec:conc}.

\section{Model}\label{sec:model}

\begin{figure}
\centering
\centerline{\epsfxsize = 12cm \epsfbox[0 53 570 744 ]{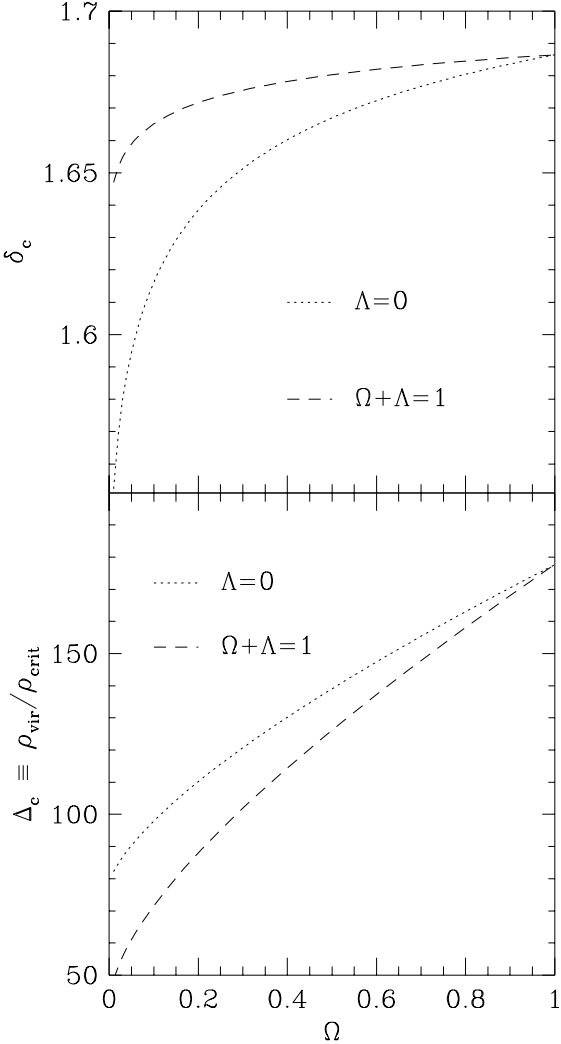}}
\caption{{\it Upper panel}: critical threshold for collapse,
$\delta_{\rm c}$, as a function of $\Omega$, in the spherical
collapse model. Results are plotted for open models with $\Lambda=0$ (dotted 
line) and flat models with $\Omega+\Lambda=1$ (dashed lines). {\it Lower 
panel}: the virial density of collapsed objects in units of the critical 
density. The dotted and dashed lines are as in the upper panel. 
}
\label{fig:scoll}
\end{figure}

An analytical expression for the comoving number density of dark
matter halos of mass $M$ in the interval $dM$, originally derived by
Press \& Schechter (1974) (see also Bond \etal 1991), is
\begin{equation}
\frac{\d n}{\d M} = \left( \frac{2}{\pi} \right)^{\frac{1}{2}}
\frac{\bar \rho}{M^2}
\frac{\dcz}{ \sigma}
\left\vert \frac{\d \ln \sigma} {\d \ln M}\right\vert
\exp \left[- \frac{\dcz^2}{2\sigma^2} \right], 
\label{ps}
\end{equation} 
where $\bar \rho$ is the present mean density of the universe, and
$\sigma(M)$ the present, linear theory {\it rms} density fluctuation in
spheres containing a mean mass $M$.  The evolution with redshift is
controlled by the redshift dependent density threshold $\dcz$.  For the
case of $\Omega_0=1$, the conventional choice of this threshold is $\dcz=
1.686 (1+z)$. This is the value, extrapolated to the present using linear
theory, of the overdensity of a uniform spherical overdense region at the
point at which the exact non-linear model predicts that it should collapse to
a singularity.  This threshold, along with the choice of top-hat filtering
to define $\sigma(M)$, gives a mass function that agrees
remarkably well with the results of N-body simulations (\eg Efstathiou
\etal 1985, Lacey \& Cole 1994).  For $\Omega \ne 1$ one can use the
spherical collapse model to derive a general expression for the density
threshold. Expressing the threshold as $\dcz = \delta_{\rm
c}(0)/D(z,\Omega_0,\Lambda_0)$ where $D(z,\Omega_0,\Lambda_0)$ is the
linear growth factor normalised to unity at $z=0$ \cite{peeb}, we find that
$\delta_{\rm c}(0)$ has only a weak dependence on $\Omega$ for both open
models with $\Lambda=0$ and flat models with $\Omega+\Lambda=1$ (see
Fig~\ref{fig:scoll}). The details of the $\Lambda=0$ calculation can be
found in Lacey \& Cole (1993; see also Maoz 1990), and the
$\Omega+\Lambda=1$ result is derived here in the Appendix (see also Lilje
1992, White \etal 1993 and Kochaneck 1995).

In order to convert the mass function obtained from (\ref{ps}) to
a temperature function we assume that the gas is isothermal. In this case, 
\begin{eqnarray}
kT_{\rm gas} &=& \frac{9.37}{\beta (5X+3)} 
\left(\frac{M}{10^{15}\Msol}\right)^{\frac{2}{3}}
 \cr
&& \times (1+z)
\left(\frac{\Omega_0}{\Omega(z)}\right)^{\frac{1}{3}}
\Delta_{\rm c}^{\frac{1}{3}} \,{\rm keV},
\label{kev}
\end{eqnarray}
where $\Delta_{\rm c}$ is the ratio of the mean halo density, to the
critical density at that redshift, $\beta$ is the ratio of specific
galaxy kinetic energy to specific gas thermal energy and $X$ is the
hydrogen mass fraction which we take to be $X=0.76$. Recent work by
Navarro \etal (1995) shows that equation~(2.2) is
accurately obeyed in N-body/hydrodynamic simulations of the formation
of clusters in universes with $\Omega_0=1$.
Their simulations predict an X-ray
luminosity-weighted $\beta=1.07 \pm 0.05$ for individual clusters, 
in fair agreement with
observational determinations (e.g. Forman \& Jones 1990). We
adopt the value $\beta=1$ throughout this paper, but our results may
be modified for different choices simply by rescaling all predicted 
temperatures by $\beta^{-1}$.

The density contrast $\Delta_{\rm c}$ is computed from the spherical
collapse model assuming that the cluster virialises at the redshift at
which we view it. Its dependence on $\Omega$ (and therefore on redshift) is
given in the lower panel of Fig~\ref{fig:scoll}. The assumption that
clusters form at the redshift at which we view them is a good approximation
in the $\Omega_0=1$ model since in this case halos are continuously
accreting material. The value $\Delta_{\rm c}=178$, appropriate to
$\Omega=1$, has recently been shown to separate well the interior of the
virialised halo from the surrounding infalling material \cite{cl}. In
low-$\Omega$ models the accretion rate onto a halo declines after $z
\approx 1/(\Omega_0^{-1} -1)$ and so little evolution in the density and
temperature of the bulk of the gas takes place at low redshift. The density
in the simple spherical collapse model, $\rho^0_{\rm crit} (1+z)^3
(\Omega_0/\Omega(z))\Delta_c$ (where $\rho^0_{\rm crit}$ is the critical
density at the present day), does not accurately reproduce this 
behaviour but the high resolution simulations needed to determine it have
yet to be carried out. Nevertheless, for the range of redshifts and
$\Omega_0$s that we consider, the simple model is a good 
approximation. 

The Sunyaev-Zel'dovich \cite{sz} effect is produced by the inverse Compton
scattering of cosmic microwave background (CMB) photons off high energy
electrons in the intracluster gas. This process distorts the blackbody
spectrum by shifting microwave photons to higher energies. At long
wavelengths (longward of $\lambda=1.37$~mm for $T_{\rm CMB} = 2.726$K),
the cluster produces a negative fluctuation in the surface brightness of
the CMB, while at shorter wavelengths it produces a positive
fluctuation. In the long wavelength regime the microwave background
decrement from the cluster is given by
\begin{equation}
\frac{\Delta T}{T}=-2y ,
\end{equation}
where $y$ is the integral of the electron pressure along a 
line-of-sight through the cluster, 
\begin{equation}
y=\int{n_{\rm e}\sigma_{\rm T}\left(\frac{kT}{m_{\rm e}c^2}\right)dl}.
\label{y}
\end{equation}
Here $n_{\rm e}$ is the number density of electrons, $m_e$ the electron
mass, and $\sigma_T$ the Thompson cross-section. We can define an
effective angular cross-section, $Y$, by integrating $y$ over the
projected area of the cluster and dividing by the square of the angular
diameter distance, $r_{\rm d}$,
\begin{equation}
Y=r_{\rm d}^{-2} \, \int{y~dA}.
\label{Y}
\end{equation}
This quantity has a simple physical interpretation. In the long wavelength
regime $2Y$ is simply the effective angular area of the microwave
background obscured by the cluster. An unresolved observation of a cluster
with a radio telescope of effective beam area, $A_{\rm beam}$,
would, in the long wavelength regime, measure a signal $\Delta T/ T=
-2Y/A_{\rm beam}$. If the cluster is resolved, then the signal depends on
the density profile of the cluster. Assuming a surface density
proportional to $R^{-1}$, as in an isothermal sphere, then $\Delta T/ T=
-2Y/(A_{\rm clus}\,A_{\rm beam})^{1/2}$, where $A_{\rm clus}$ is the
angular cross-section of the virialised cluster. Using the virial
radius to define the edge of the cluster, we can then write
\begin{eqnarray}
A_{\rm clus}&=&~64.1 ~\beta (5X+3) \left(\frac{kT_{\rm gas}}{\rm keV}\right) 
(1+z)^{-3} \left(\frac{\Omega(z)}{\Omega_0}\right) \cr
&& \times \left(\frac{\Delta_{\rm c}}{178}\right)^{-1}
\left(\frac{r_{\rm d}}{100 \Mpc}\right)^{-2} {\rm arcmin}^2.
\end{eqnarray}

The effective angular area, $Y$, is also easily related to the CMB flux
absorbed (or re-emitted) by the cluster at other frequencies. Defining a
dimensionless frequency $ x= {\rm h_{p}} \nu / k T_{\rm CMB} =
\lambda_0/\lambda$, where $h_p$ is Planck's constant and
$\lambda_0=5.28$mm for $T_{\rm CMB} = 2.726$K \cite{tcmb}, the
emitted flux is
\begin{eqnarray}
S_\nu(x) &=& S_\nu^{\rm CMB}(x) \,\, Q(x) \,\, Y\\
&=& 2.29\times 10^{4}  \,  \frac{x^3}{e^x-1} \,  Q(x)  \, 
\left(\frac{Y}{\rm arcmin^2} \right)  \, {\rm mJy} ,
\end{eqnarray}
where
\begin{equation}
Q(x) = \frac{x e^x}{e^x-1} \left[ \frac{x}{\tanh(x/2)}-4\right] .
\end{equation}
and 1~mJy$\equiv 10^{-29}$ Js$^{-1}$m$^{-2}$ Hz$^{-1}$.
In the long wavelength limit, $x \rightarrow 0$, $Q(x) \rightarrow -2$
and this reduces to the result stated earlier,
$\Delta T/ T = S_\nu / S_\nu^{\rm CMB} = -2Y$. At $x\approx3.83$
($\lambda=1.37$mm) $Q(x)=0$ and at higher frequencies the cluster
appears as a source of emission.

\begin{figure*}
\centering
\centerline{\epsfysize = 10cm \epsfbox[20 250 580 600]{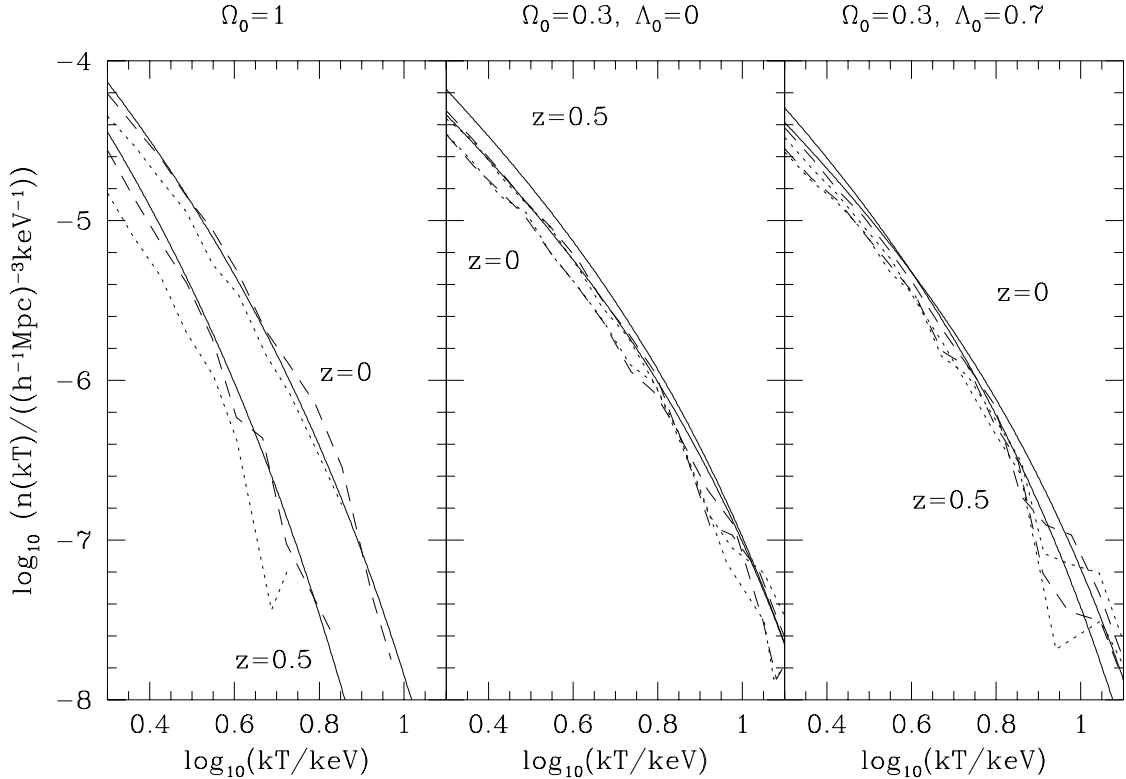}}
\caption{Comparison of the temperature functions predicted by the
Press-Schechter distribution and the results of N-body simulations. 
For each of the three cosmological models (with 
parameters given at the top of each panel), temperature functions are
plotted for $z=0$ and $z=0.5$. The Press-Schechter predictions, which 
are normalised according to the spherical collapse model, are shown 
by solid lines. The simulation results are plotted with dashed lines for
clusters identified with a friends-of-friends algorithm and with dotted
lines for clusters found with the spherical overdensity algorithm. Over
the full range of abundances and redshifts shown the agreement between
theory and simulations is very good.
}
\label{fig:nbody}
\end{figure*}

The effective cross-section, $Y$, defined by equations (\ref{y}) and
(\ref{Y}), is proportional to the total mass and average temperature of
the intracluster gas. Specifically, for an isothermal intracluster gas,
\begin{eqnarray}
Y &=& \frac{\sigma_{\rm T}}{2 m_{\rm e} m_{\rm p} c^2} \,
      f_{\rm ICM} \, (1+X) \,\, M \, kT_{\rm gas} \,  
r_{\rm d}^{-2} \cr \nonumber
&=& 9.68\times10^{-2}{\rm h} \, f_{\rm ICM} \, (1+X) \, \,
\left(\frac{M}{10^{15} \Msol}\right) \\ 
&&\times \left(\frac{kT_{\rm gas}}{\rm keV}\right) \,
\left(\frac{r_{\rm d}}{100 \Mpc}\right)^{-2}
~{\rm arcmin}^2,
\label{yofm}
\end{eqnarray}
where $f_{\rm ICM}$ is the fraction of the cluster mass represented by the
hot intracluster gas. We adopt $\beta=1$ and 
$f_{\rm ICM}=0.1$ but our results can readily be rescaled to other choices.
Note that here it is appropriate to use the mass-weighted
temperature for which  Navarro \etal (1995) find $\beta=1.2$
for their $\Omega_0=1$ simulations.

\section{Comparison with N-body Simulations}\label{sec:sims}

In order to assess the accuracy of the Press-Schechter mass distribution
in the regime of interest -- the mass scale of rich galaxy clusters -- we
compare the model predictions with the abundance of dark matter
clumps found in a new set of large cosmological N-body simulations (Cole,
Frenk \& Weinberg in preparation). The simulations were performed with 
the AP$^3$M code of Couchman \shortcite{ap3m} using $192^3\approx 7 \times
10^6$ particles in a periodic box of size $l_{\rm box}=345.6 \Mpc$. The
particle mass was $M_{\rm p}=1.64\times 10^{12} \Omega_0 \Msol$ and the
force resolution $\epsilon=180 \kpc$, where $\epsilon$ is the equivalent 
Plummer potential softening parameter. Two sequences of
simulations were carried out with different values of $\Omega_0$, a 
sequence of open models with $\Lambda_0=0$ and a sequence 
of flat models with $\Omega_0+\Lambda_0=1$. Both were
normalised to have $\sigma_8= 0.55 \, \Omega_0^{-0.6}$ 
so as to reproduce,
approximately, the observed abundance of galaxy clusters (White \etal 1993). 
The number of timesteps required to evolve the simulations accurately from the
linear regime to the present epoch was approximately $100/\Omega_0$. The
initial linear power spectrum was the same in all simulations, a CDM-like
spectrum with scale parameter $\Gamma=\Omega_0 {\rm h}=0.25$. This value of
$\Gamma$ is suggested by observations of large-scale galaxy clustering (eg.
Maddox \etal 1990). Here we will consider only three representative
models, one with $\Omega_0=1$, another with $\Omega_0=0.3$ and
$\Lambda_0=0$, and a third one with $\Omega_0=0.3$ and $\Lambda_0=0.7$.

We identified groups of particles in the simulations using two different
algorithms. The first was the standard friends-of-friends algorithm (Davis
\etal 1985) with linking length $b_l$ times the mean interparticle
separation; the second was the spherical overdensity algorithm \cite{lc2}
with density contrast $\kappa_{\rho}$. We chose values of $b_l$ and 
$\kappa_{\rho}$ so
that, on average, groups have the overdensity characteristic of virialised
objects predicted by the spherical collapse model. For the
friends-of-friends linking length we took $b_l=0.2$ in the $\Omega_0=1$
simulation and scaled as $b_l \propto (\Delta_{\rm 
c}/\Omega)^{-1/3}$ for the other cases. In both the simulations and the
Press-Schechter model we converted mass to temperature using
relation~(\ref{kev}).

Fig~\ref{fig:nbody} compares cluster abundances as a function of $kT$
at $z=0$ and $z=0.5$ in the N-body simulations and in the Press-Schechter
model. Overall, the analytical predictions are in excellent agreement with
the N-body results and reproduce the near exponential fall-off of the
temperature functions very accurately. The threshold density, $\delta_c$,
has {\it not} been treated as a free parameter, but has instead been fixed
at the value prescribed by the spherical collapse model. For the
$\Omega_0=1$ case at $z=0$ this value is essentially perfect and produces
a temperature function midway between that obtained using the
friends-of-friends and spherical overdensity algorithms. For $z=0.5$, the
spherical overdensity algorithm finds somewhat fewer high temperature
clusters than either the friends-of-friends algorithm or the
Press-Schechter model. This may partially reflect the relatively low
resolution of the N-body simulations as the most massive groups contain
only about $100$ particles in the $\Omega_0=1$ simulation at $z=0.5$. In
the two $\Omega_0=0.3$ cases, the Press-Schechter model matches the very
slow evolution of the halo abundances in the simulations extremely well.
If one were to adjust $\delta_{\rm c}$ upwards by approximately 4\%, then
the N-body results would be reproduced almost perfectly over the
temperature range plotted in Fig~\ref{fig:nbody}. However, such a small
adjustment is negligible compared to the uncertainty in the
abundance of real clusters. We note that for the more common, lower
temperature clusters not shown in this figure, the Press-Schechter
distribution predicts a significantly higher abundance than found in the
simulations. We conclude that, for the range of masses of interest, the 
agreement with the numerical results is sufficiently good that we can 
confidently use the Press-Schechter model for detailed calculations. 

\begin{figure*}
\centering
\centerline{\epsfbox[10 225 580 580]{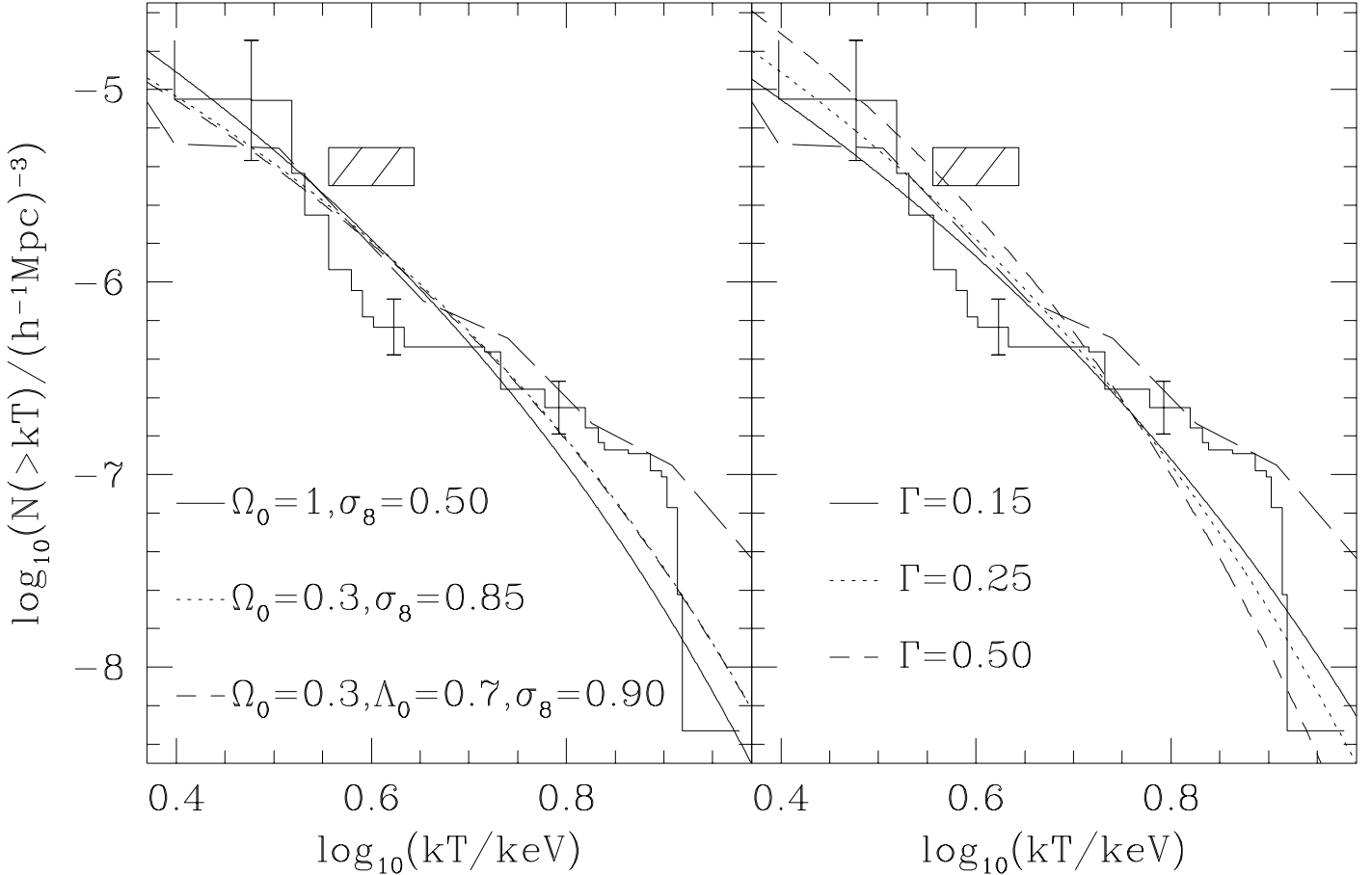}}
\caption{Predicted and observed X-ray temperature functions. The left-hand
panel shows model predictions for $\Gamma=0.25$ and $\Omega_0=1$ (smooth
solid line), $\Omega_0=0.3$, $\Lambda_0=0$ (dotted line) and
$\Omega_0=0.3$, $\Lambda_0=0.7$ (dashed line). The models are normalised by
fitting to the observed data as described in the text. The observed
temperature function, shown as a solid line with steps, was derived from
the data compiled by Henry \& Arnaud (1991). The error bars at the three
temperatures where the models were fit were obtained from a bootstrap
analysis. The best-fit values of $\sigma_8$ are given in the figure. The
long-dashed line shows the
cumulative temperature function obtained by Edge {\it et al} (1990). The
hatched box represents the range of normalisations derived by White,
Efstathiou \& Frenk (1993) from a similar theoretical analysis of combined
X-ray and optical data. The right-hand panel gives model predictions
for $\Omega=1$, $\sigma_8=0.5$ and three values of $\Gamma$: 0.15 (solid
line), 0.25 (dotted line), and 0.50 (dashed line). The observational data
and the hatched box are the same as in the left-hand panel.
}
\label{fig:ctfnz0}
\end{figure*}

\section{Normalisation of the models}\label{sec:norm}

We now compare the predicted and observed distributions of cluster X-ray
temperatures in the local universe in order to obtain an estimate of
$\sigma_8$, the {\it rms} mass fluctuation in spheres of radius $8 \Mpc$.
The observational data we use is the complete flux-limited sample of
25~clusters compiled by Henry \& Arnaud \shortcite{ha}. We compare our
estimate with earlier determinations of $\sigma_8$ from both X-ray and
optical data (Henry \& Arnaud 1991; White \etal 1993; Viana \& Liddle
1995) and discuss the reasons why these earlier determinations gave
slightly different values of $\sigma_8$.

The stepped curve in Fig~\ref{fig:ctfnz0} shows the following simple
estimate of the cumulative cluster temperature function derived from the 
Henry \& Arnaud data, 
\begin{equation}
N(>kT) = \sum_{kT_i>kT} 1/V_{\rm max,i}, 
\label{esti}
\end{equation}
where $V_{\rm max,i}$ is the maximum volume in which the cluster could be
detected given the flux limit and geometric boundaries of the survey. We
have chosen to present the results in cumulative form in order to avoid
binning the data. Since the cluster abundance falls very rapidly with
increasing temperature, the differential temperature function averaged
over each bin can be significantly larger than the underlying unbinned
distribution. Henry \& Arnaud (1991) estimated the differential
temperature function by weighting the number of clusters in each
temperature bin with the {\it average} value of $V_{\rm max,i}$ in that
bin. This estimator is only equivalent to (\ref{esti}) if $V_{\rm max,i}$
is the same for all of the clusters in each bin. However, since the bins
have non-zero width and the $L_{\rm X}$--$T_{\rm X}$ relation has
considerable scatter, there is also considerable scatter in the individual
$V_{\rm max,i}$ values in each temperature bin.
For the Henry \& Arnaud dataset this is a large effect and
would have led them to underestimate the cluster abundance by
approximately a factor of $4$. However, (Henry private communication) in
computing each $V_{\rm max,i}$, a spurious factor of $4.2$ entered their
calculation and this largely compensated for the bias in the estimator.
Thus the published Henry \& Arnaud (1991) temperature function is in
reasonably good agreement with our estimate. It is also in reasonable
agreement with the cumulative temperature function of Edge \etal
\shortcite{edge}, reproduced as the long dashed line in 
Fig~\ref{fig:ctfnz0}. 

To estimate $\sigma_8$, we fit our model predictions to our estimate of 
the temperature function. First, we calculate the
statistical errors in the temperature function through a boostrap
procedure. At three selected temperatures, $T_i$, we computed (using
equation~4.1) the temperature function for a large number of samples of
25 clusters
constructed by randomly selecting, with replacement, from the 
original list of 25 clusters in Henry \& 
Arnaud's compilation. The error bars in Fig.~\ref{fig:ctfnz0} show the
resulting $1-\sigma$ ranges in the bootstrap distribution of $\log_{10} N_{\rm
boot}(>kT)$, at our three chosen temperatures. This distribution was also
used to compute the covariance between the estimates at different
temperatures. Averaging over the bootstrap samples we thus obtain the
covariance matrix, 
\begin{equation}
    C_{ij} = \langle \epsilon_i \epsilon_j \rangle,
\end{equation}
where 
$\epsilon_i = \log_{10} N_{\rm boot}(>kT_i) - \log_{10} N_{\rm data}(>kT_i) $.
We then fit the model temperature functions by minimising 
\begin{equation}
\chi^2 = \sum_{ij} \delta_i  C^{-1}_{ij} \delta_j,
\end{equation}
where 
$ \delta_i = \log_{10} N_{\rm model}(>kT_i) -\log_{10} N_{\rm data}(>kT_i) $.
If the data points are uncorrelated $C_{ij}$ is diagonal and
equation~(4.3) reduces to the normal definition of $\chi^2$.  We find that
there are significant correlations between the estimates of $\log_{10}
N(>kT)$ at the three selected temperatures, but that the models which
minimise $\chi^2$ are insensitive to whether the correlations are treated
as above or simply ignored.

The best-fit models are shown in Fig~\ref{fig:ctfnz0} for $\Omega_0=1$
and $\Omega_0=0.3$, with and without a cosmological constant. The models in
the left-hand panel all have the same CDM-like power spectrum with
$\Gamma=0.25$. Because of the very sensitive dependence of cluster
abundance on spectrum normalisation, $\sigma_8$ can be estimated with high
precision even though the errors in the empirical cluster abundance
are quite large. The formal error on $\sigma_8$ from the fits in Fig~3
is $\pm 2\%$, but this is likely to be an underestimate of the true
uncertainty because, with only 25~clusters, the errors in the temperature
function are unlikely to be Gaussian and, in addition, systematic errors
are likely to be significant. 

The right-hand panel of Fig~3 shows the effect of allowing $\Gamma$ to
vary for the case of $\Omega_0=1$. Values of $\Gamma<0.25$ produce more
large scale power and more very hot clusters.  This produces a temperature
function with a shallower slope, which is in better agreement with the
observations. (The best fitting model has $\Gamma=0.03$ and $\sigma_8=0.51$,
but this is not a
reliable way of constraining $\Gamma$ because the high and low temperature
points are anticorrelated, and this creates a large uncertainty in the
slope of the temperature function.) Fortunately, the best fitting values of
$\sigma_8$ are quite insensitive to the adopted value of $\Gamma$, because
as $\Gamma$ varies it is the slope of the temperature function that varies
with the pivot point remaining in the middle of the range of rich cluster
temperatures.  This simply reflects the fact noted above that the average
mass within a sphere of radius $8 \Mpc$ in an unperturbed universe with
$\Omega_0 \approx 1$ is very close to the mass of a rich galaxy cluster.

In summary, we find that the observed temperature function is well fit by 
all our CDM-like models if 
\begin{equation}
\sigma_8 = (0.50 \pm 0.01) \Omega_0^{-0.47+0.10\Omega_0} \ \ {\rm for} 
\ \ \Lambda_0=0
\label{sig8nolam}
\end{equation}
and
\begin{equation}
\sigma_8 = (0.50 \pm 0.01) \Omega_0^{-0.53+0.13\Omega_0} \ \
{\rm for} \ \ \Omega_0+\Lambda_0=1. 
\label{sig8lam}
\end{equation}
The quoted statistical uncertainty corresponds to the standard
deviation of the $\sigma_8$ values obtained from fitting the
individual bootstrap catalogues. A more realistic estimate of the
error would take into account the variation of $\sigma_8$ due to the
uncertainty in the measured X-ray temperatures and the values of both
$\delta_{\rm c}$ and $\beta$.  The error in $\delta_c$ we estimate
from Fig.~2 to be $\lsim 4 \%$. The uncertainty in the appropriate
$\beta$ we take to be $10\%$ (see Navarro \etal 1995).  Note that
adopting $\beta=1.1$ leads to values of $\sigma_8$ which are $6\%$
larger than those quoted above.  Combining these sources of error we
estimate the overall uncertainty in $\sigma_8$ to be approximately
$8\%$, which is four times as large as the statistical error shown in
equations (4.4) and (4.5).

The values of $\sigma_8$ that we infer from the X-ray data are
systematically lower than those obtained by White \etal (1993) who 
found $\sigma_8 = 0.57 \Omega_0^{-0.56}$.  Our modelling of
the cluster abundances is almost identical and the difference in our
inferred values of $\sigma_8$ results almost entirely from the different
observational data that we fitted.  White \etal estimated the mass of
clusters with abundance $4\times 10^{-6} {\rm h^3 Mpc^{-3} }$ in two
different ways. The first and larger estimate was based on the median
velocity dispersion of rich Abell clusters. The second was based on the
cumulative temperature functions of Henry \& Arnaud \shortcite{ha} and
Edge \etal \shortcite{edge}. The range spanned by these two estimates,
expressed as a temperature rather than a mass using equation~(2.2), is
indicated by the hatched box in Fig~\ref{fig:ctfnz0}. The lower estimate
of the X-ray temperature is slightly higher than our analysis of the Henry
\& Arnaud data implies. This difference is due to the fact that the
White \etal estimate was inferred indirectly from the differential
temperature function presented by Henry \& Arnaud, not directly from the
cumulative temperature function. (If we adopt $\sigma_8 = 0.57
\Omega_0^{-0.56}$ then our model temperature curves pass through the
shaded box in agreement with White \etal).

Henry \& Arnaud \shortcite{ha} found $\sigma_8=0.59 \pm 0.02$ for
$\Omega_0 =1$ and a power-law fluctuation spectrum. They used
essentially the same data as us, but they adopted $\beta=1.2$ in
equation~(\ref{kev}), while we have assumed 
$\beta=1$. Thus, for the same value of $\sigma_8$, their model temperature
functions are shifted to lower temperatures by $\sim 0.08$ in $\log_{10}(kT)$,
and this largely accounts for the difference in the inferred values of 
$\sigma_8$. Viana
\& Liddle \shortcite{vl} obtained $\sigma_8=0.6$ for $\Omega_0=1$, with a
dependence on $\Omega_0$ close to that which we find, by fitting only to
the abundance of clusters at $7$keV. It may be seen from
Fig~\ref{fig:ctfnz0} that fitting the cumulative temperature function just
at this temperature yields a higher value of $\sigma_8$ than the one
obtained from fitting all the X-ray data.  Given the small number of
clusters with temperatures as great as $7$keV, fitting over a wider
temperature range seems more appropriate. 

In summary, our modelling of the cluster temperature function is consistent
with those of Henry \& Arnaud \shortcite{ha}, White \etal \shortcite{s8}
and Viana \& Liddle \shortcite{vl}. The range in the values of $\sigma_8$
deduced by these authors arises from the different data points they chose
to fit and, in the case of Henry \& Arnaud \shortcite{ha}, from the value
of $\beta$ they used to relate virial to gas temperature. Assuming
$\beta=1$, we conclude that the X-ray data are best fit by the values of
$\sigma_8$ given by equations (\ref{sig8nolam}) and (\ref{sig8lam}). Since
the abundance of Abell clusters of richness class $\ge 1$ is $8\times
10^{-6} {\rm h^3 Mpc^{-3} }$, our results imply that the median 1-D virial
velocity dispersion of these clusters should be approximately $650
\kms$. This is smaller than the median values of around $800 \kms$, found
in the compilations of Zabludoff, Huchra \& Geller
(1990) and Girardi \etal (1993). These differences may be understood if, as
argued by Frenk \etal (1990) and others, the higher velocity dispersion
estimates are artificially boosted by contamination from infalling groups
around the cluster. Alternatively, these larger velocity dispersions (and a
higher value of $\sigma_8$) would be consistent with the X-ray data if the
intracluster gas were significantly cooler than the virial temperature, but
this would require $\beta \gsim 1.5$. Such large values of $\beta$ are not
supported either by the data or by recent hydrodynamic simulations of
cluster formation (Navarro \etal 1995, Evrard \etal 1995).
\begin{figure}
\centering
\centerline{\epsfxsize = 8.0cm \epsfbox[50 100 574 724]{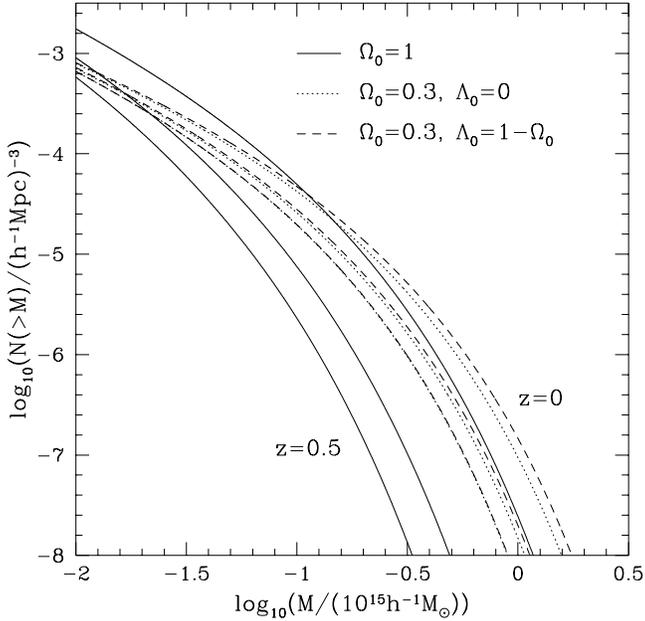}}
\caption{Predicted evolution of the cluster mass function. The 
comoving number density of clusters per $({\rm h^{-1} Mpc})^3$
with masses larger than $M$ is shown as a function of $M$.
Solid lines
correspond to $\Omega=1$; dotted lines to an open model with
$\Omega_0=0.3$; and the dashed lines to a flat model with $\Omega_0=0.3$ and 
$\Lambda_0=0.7$. Predictions for $z=0$, $z=0.33$ and $z=0.5$ are plotted.
There is relatively little evolution in the $\Omega_0<1$ cosmologies but, in an
$\Omega_0=1$ model, the abundance of clusters declines precipitously with
redshift.}
\label{fig:mevol}
\end{figure}

\begin{figure}
\centering
\centerline{\epsfxsize =11.5cm \epsfbox[0 0 574 724]{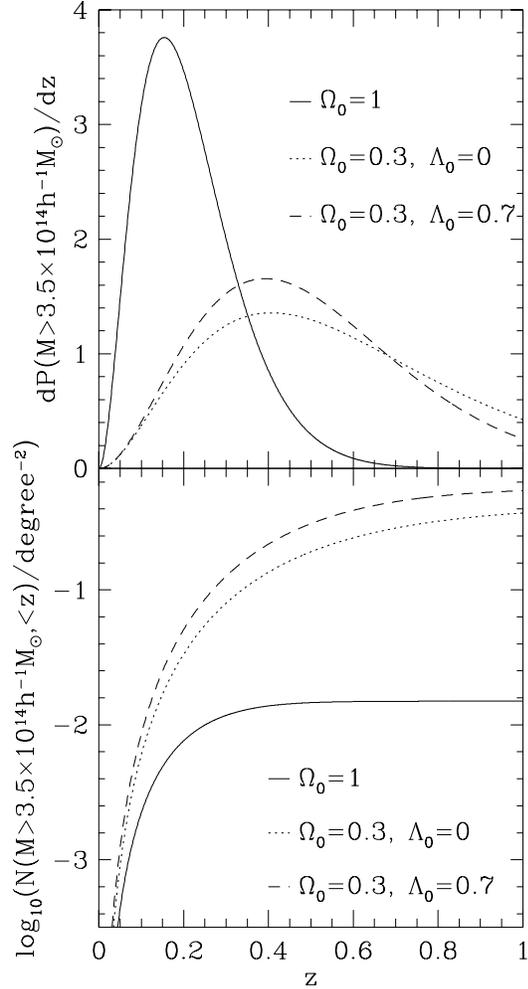}}
\caption{{\it Upper panel:} Redshift distribution of massive clusters ($M
> 3.5 \times 10^{14} \Msol$) in different cosmological models. The
ordinate gives the 
probability distribution of clusters per unit redshift interval. {\it
Lower panel:} Number counts of clusters with mass $M> 3.5 \times 10^{14}
\Msol$ out to a given redshift. The ordinate gives the count per unit area
on the sky. In both panels solid lines correspond to $\Omega_0=1$, dotted
lines to $\Omega_0=0.3$ and dashed lines to $\Omega_0=0.3, \Lambda_0=0.7$.
The models are normalised by the value of $\sigma_8$ for which the
predicted temperature function at $z=0$ best fits the data. The
low-$\Omega_0$ cosmologies produce significantly more clusters at high
redshifts than the $\Omega_0=1$ model. 
}
\label{fig:mclusz}
\end{figure}

\section{Cluster Evolution}\label{sec:evol}

Having fixed the normalisation of the models by requiring that they should 
match the local abundance of rich clusters, we now consider their
evolutionary properties. Specifically, we calculate the redshift dependence
of the mass function, X-ray temperature function and the distribution of
S-Z decrements. For each model, the evolution of the cluster mass function
(equation~(\ref{ps}) is governed by the linear growth factor,
$D(z,\Omega_0,\Lambda_0)$. The evolution of the X-ray temperature and S-Z
effect depends, in addition, on the evolution of the virial density
$\Delta_{\rm c} \Omega_0/\Omega(z)$, which determines the  relation
(\ref{kev}) between mass and temperature.

\subsection{Evolution of the cluster mass function}\label{sec:mevol}

The growth of fluctuations continues at a rapid rate at recent times if
$\Omega_0=1$ and very little if $\Omega_0=0.3$. As a result, 
the rich cluster mass function evolves dramatically between $z=0.5$ and
$z=0$ if $\Omega_0=1$, 
but much less so if $\Omega_0$ is low, whether or not the
cosmological constant is zero. This evolution is illustrated in
Fig~\ref{fig:mevol}. For $\Omega_0=1$, the comoving number density of
clusters of virial mass $M=3.5 \times 10^{14} \Msol$, typical of Abell clusters
of richness class $R\ge1$, declines by a factor of 30 between $z=0$ and
$z=0.33$ and, by $z=0.5$, it is tiny. By contrast, for $\Omega_0=0.3$,
$\Lambda_0=0.7$, the abundance of clusters of this mass has only dropped by
a factor of $\sim 7$ below the present day abundance even at $z=0.5$; if
$\Lambda_0=0$ the decline is even slower.

The strong $\Omega_0$-dependence of the rate at which the mass function
evolves is reflected in the expected redshift distributions of massive
clusters ($M>3.5 \times 10^{14} \Msol$), illustrated in the
top panel of Fig~\ref{fig:mclusz}. For $\Omega_0=1$ the distribution
peaks sharply at very low redshift, whereas for low $\Omega_0$ a broader
peak is displaced to higher redshift. The effect of $\Lambda_0$ is to move
the peak back to a somewhat lower redshift, reflecting the slightly later
epoch at which structure ceases to grow in non-zero $\Lambda_0$ cosmologies.

Integrating over the redshift distributions yields the number count of
clusters per unit area on the sky. The change in the volume element
corresponding to a fixed redshift interval enhances the differences
between the high and low-$\Omega$ models. At redshift $z=0.33$, the
volumes per unit redshift are in the ratios $1:1.23:1.7$ for $\Omega_0=1$,
$\Omega_0=0.3$, $\Lambda_0=0$ and $\Omega_0=0.3$, $\Lambda_0=0.7$
respectively. At $z=0.5$ the corresponding ratios increase to
$1:1.35:2.02$. Thus, for $\Omega_0=1$, we expect to find only $0.015$ clusters
per square degree with mass greater than $3.5 \times 10^{14} \Msol$ out to
$z=0.5$ and virtually none at higher redshifts. By contrast, for
$\Omega_0=0.3$, we expect to find more than 10 times as many clusters above
this mass with $z<0.5$. Note that predictions for low-$\Omega$ models are
relatively insensitive to the value of $\Lambda_0$, with
only about $50\%$ more clusters predicted to exist 
in the non-zero $\Lambda_0$ model.
This factor results largely from the difference in the volume elements.

The cluster mass that enters into equation~(2.1) and
Figs~\ref{fig:mevol} and~\ref{fig:mclusz} is the virial mass, {\it ie}
the mass contained within a sphere of mean overdensity $\Delta_c$. 
In practice, gravitational lensing measurements give the
mean projected surface density, $\Sigma(R)$, within a radius, $R$, which is
typically less than $1 \Mpc$. To relate these two masses requires a model
of the cluster mass distribution. Recent high resolution N-body
simulations show that, over most of the cluster, the dark matter density
profile is well approximated by an isothermal profile (Navarro \etal
1995; Cole \& Lacey 1996). Although in principle it would be
straightforward to use the analytical fit to the N-body profile given by
Navarro \etal (1995), for our purposes the isothermal profile 
approximation is quite adequate. In this case, mean 
surface density is related to virial mass by 

\begin{equation}
M_{\rm vir}= {4\over H} \left( {GR^3 \Sigma(R)^3 \over \Delta_c}\right)^{1/2},
\label{virm}
\end{equation}
where $R$ is any radius inside the virial radius, $H$ is Hubble's constant 
at the redshift of the cluster and 
the overdensity $\Delta_c$ is given in Fig~1 as a function of
$\Omega$. This formula may be used to relate the surface mass
density estimated from weak gravitational lensing analyses to the virial
mass used throughout this paper. Note, however, that we have implicitly
assumed that any foreground or background mass makes a negligible
contribution to the lensing signal.

\begin{figure}
\centering
\centerline{\epsfxsize = 8.0cm \epsfbox[50 100 574 724]{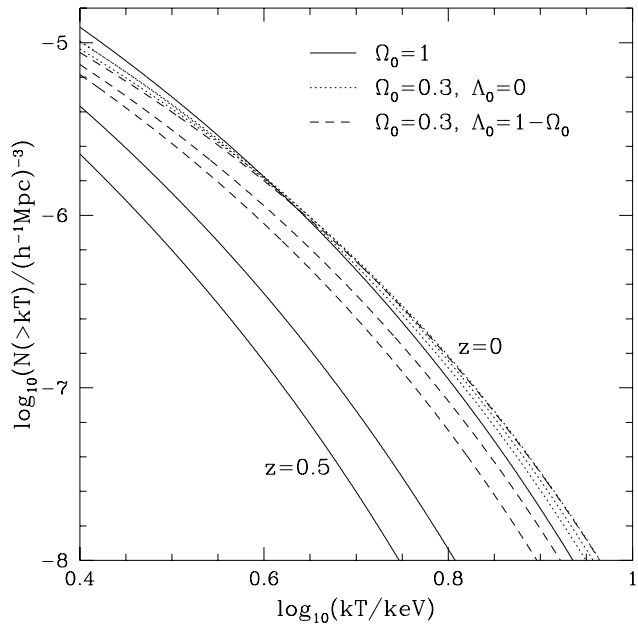}}
\caption{Predicted evolution of the cluster X-ray temperature function. The
comoving number density of clusters per $({\rm h^{-1} Mpc})^3$
hotter than $kT$ is shown as a function of $kT$.
Solid lines
correspond to $\Omega=1$; dotted lines to an open model with
$\Omega_0=0.3$; and the dashed lines to a flat model with $\Omega_0=0.3$ and 
$\Lambda_0=0.7$. Predictions for $z=0$, $z=0.33$ and $z=0.5$ are plotted.
There is little evolution in the $\Omega_0<1$ cosmologies but in an
$\Omega_0=1$ model, the abundance of clusters declines precipitously with
redshift.}
\label{fig:tevol}
\end{figure}

\begin{figure}
\centering
\centerline{\epsfxsize =11.5cm \epsfbox[0 0 574 724]{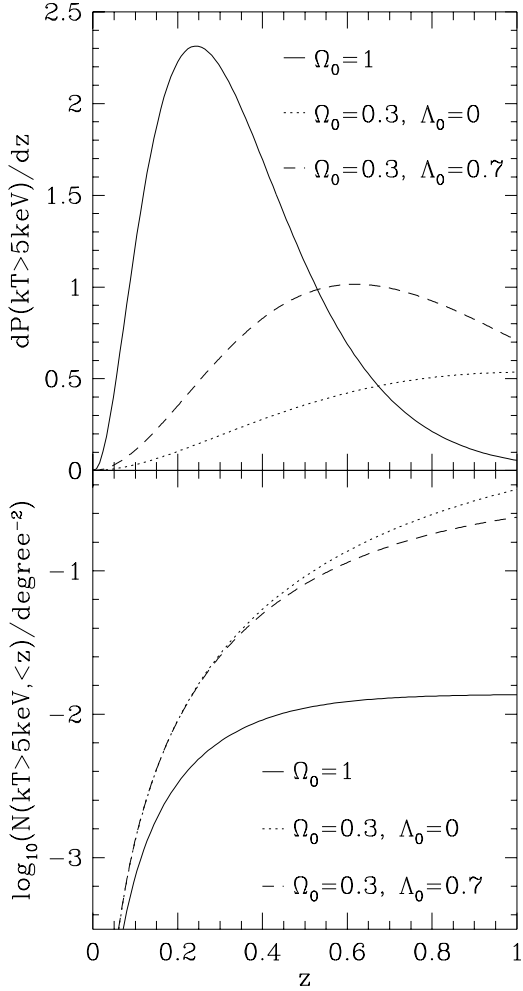}}
\caption{{\it Upper panel:} Redshift distribution of hot clusters ($kT > 5
{\rm keV}$) in different cosmological models. The ordinate gives the
probability distribution of 
clusters per unit redshift interval. {\it Lower panel:} Number
counts of clusters hotter with $kT > 5 {\rm keV}$ out to a given redshift.
The ordinate gives the count per unit area on the sky. In both panels
solid lines correspond to $\Omega_0=1$, dotted lines to $\Omega_0=0.3$ and
dashed lines to $\Omega_0=0.3, \Lambda_0=0.7$. The models are normalised
by the value of $\sigma_8$ for which the predicted temperature function at
$z=0$ best fits the data. The low-$\Omega_0$ cosmologies produce
significantly more clusters at high redshifts than the $\Omega_0=1$ model.
}
\label{fig:nclustz}
\end{figure}

\begin{figure}
\centering
\centerline{\epsfxsize = 8.0cm \epsfbox[50 100 574 724]{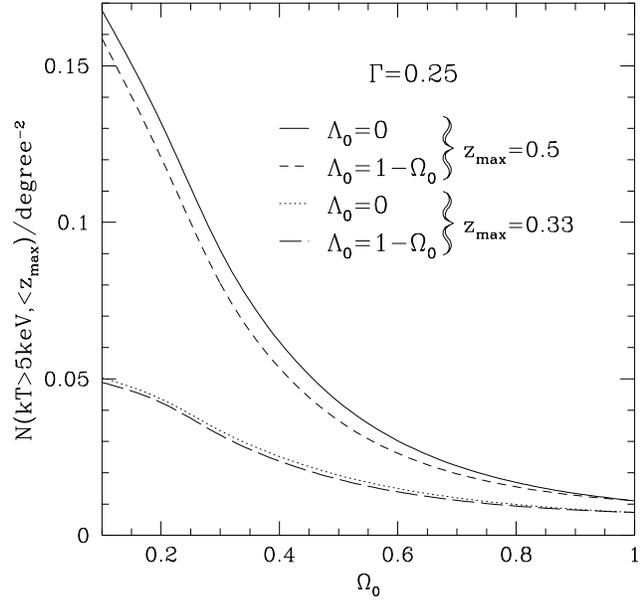}}
\caption{Predicted number counts of clusters hotter than $5$ keV out to
$z=0.33$ and $0.5$, as a function of $\Omega_0$. The solid and dotted lines
correspond to open models ($\Lambda_0=0$) and the dashed lines to flat
models ($\Lambda_0=1-\Omega_0$). The abundances are sensitive to the
choice of $\sigma_8$, but the ratio of the numbers expected in different
cosmologies is less so.
}
\label{fig:ntzom}
\end{figure}

\subsection{Evolution of the X-ray temperature function}\label{sec:tevol}

The evolution of the X-ray temperature function depends both on the growth
factor, $D(z,\Omega_0,\Lambda_0)$, and on the virial density, $\Delta_{\rm
c} \Omega_0/\Omega(z)$. Thus, at high $z$, the abundance of clusters of a
given temperature is determined by the balance between the overall decline
in the population of virialized clusters and the lower mass associated
with each temperature. If $\Omega_0=1$, the first factor is dominant and the
temperature function declines precipitously (see Fig~\ref{fig:tevol}).
By contrast, in an open, $\Omega_0=0.3$, universe, the two effects nearly
cancel out and the temperature function remains virtually unchanged at
least out to $z=0.5$. A flat, $\Omega_0=0.3$, universe is intermediate and,
in this case, the temperature function declines slowly with redshift.

The redshift distribution of clusters hotter than 5 keV is shown in
Fig~\ref{fig:nclustz}. As was the case for the mass function, the
$\Omega_0=1$ model produces large clusters predominantly at low redshifts.
The $\Omega_0=0.3$ models, on the other hand, give rise to extended
redshift distributions. Again, the lower redshift at which structure
``freezes-out" when a $\Lambda_0$ term is included produces somewhat
stronger evolution in this case compared to an open cosmology. When
integrating over redshift, the effect of different evolutionary rates is
enhanced by the larger volume in low density universes. As a result, the
number counts per unit area of sky, displayed in the lower panel of
Fig~\ref{fig:nclustz}, depend strongly on the value of $\Omega_0$. For
our choice of parameters, the counts in the two $\Omega_0=0.3$ models are
very similar and the total number of clusters hotter than 5 keV at
redshifts less than 0.5 is about 9 times higher in these models
than in the $\Omega_0=1$ case. The expected number counts for different
values of $\Omega_0$ are shown in Fig~\ref{fig:ntzom}, for both open
and flat models.

We emphasize that the results shown in
Figs~\ref{fig:tevol}~-~\ref{fig:ntzom} are very sensitive to the
normalisation of the fluctuation amplitude. Our adopted values of
$\sigma_8$ were fixed by requiring that each model should agree well with
the present day X-ray temperature function. With this particular choice,
the inclusion of a $\Lambda_0$ term turns out to make very little
difference to the predicted abundance of hot clusters out to redshift of
0.5. However, the number counts out to this redshift do discriminate well
between different values of $\Omega_0$.

\subsection{Evolution of Sunyaev-Zel'dovich effect}\label{sec:yevol}

\begin{figure}
\centering
\centerline{\epsfxsize = 8.0cm \epsfbox[50 100 574 724]{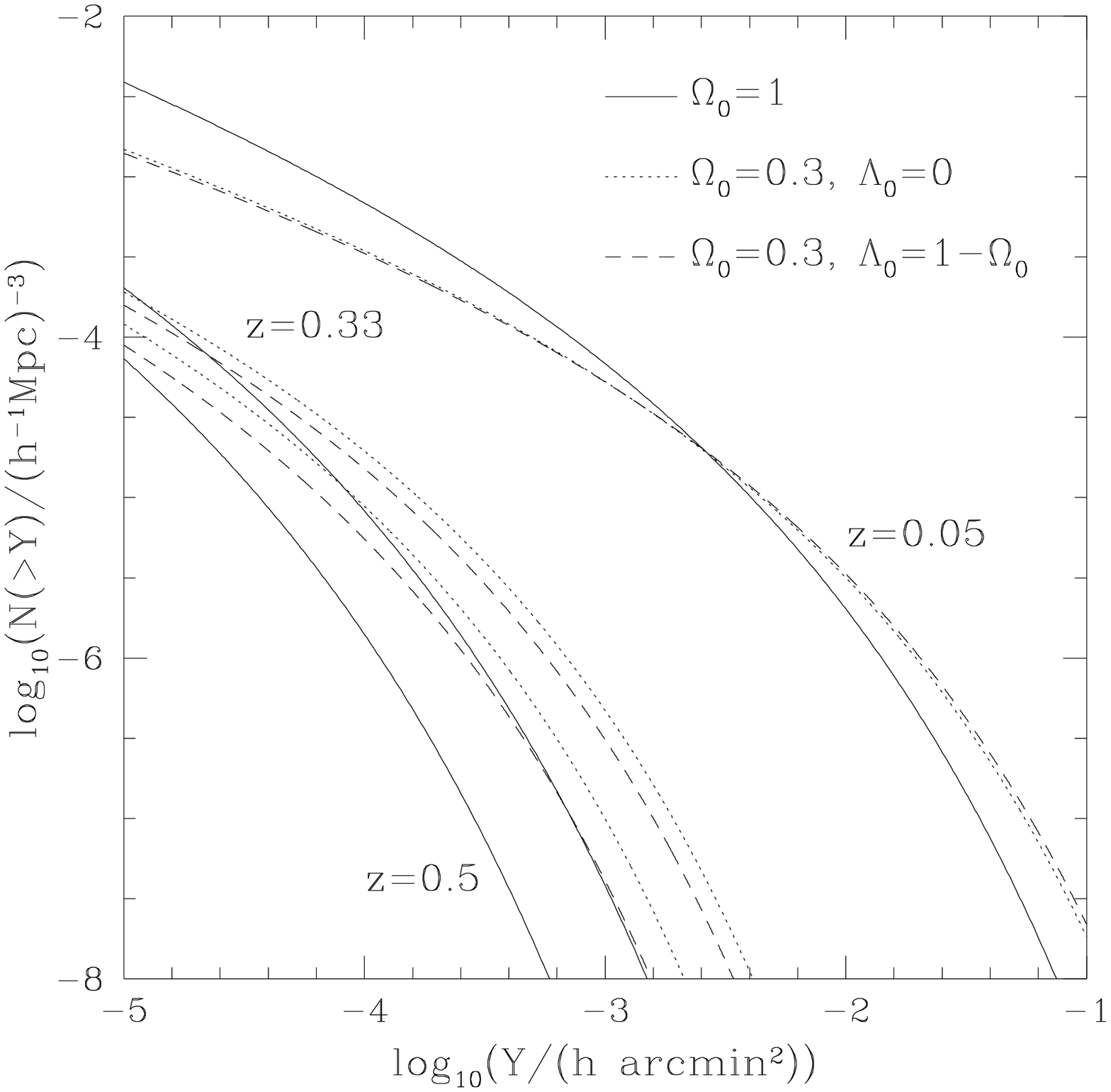}}
\caption{Predicted evolution of the cumulative cluster S-Z $Y$-function.
The ordinate gives the comoving number density of clusters per $({\rm
h^{-1} Mpc})^3$. The solid lines are for $\Omega_0=1$; the dotted lines
for $\Omega_0=0.3$ and the dashed lines for $\Omega_0=0.3$ and $\Lambda_0=0.7$.
Predictions for $z=0.05$, $z=0.33$ and $z=0.5$ are plotted. Yhe
$Y$-function evolves significantly in the three cosmologies, but the
evolution is strongest for $\Omega_0=1$.
}
\label{fig:yevol}
\end{figure}

\begin{figure}
\centering
\centerline{\epsfxsize =8.0cm \epsfbox[50 100 574 724]{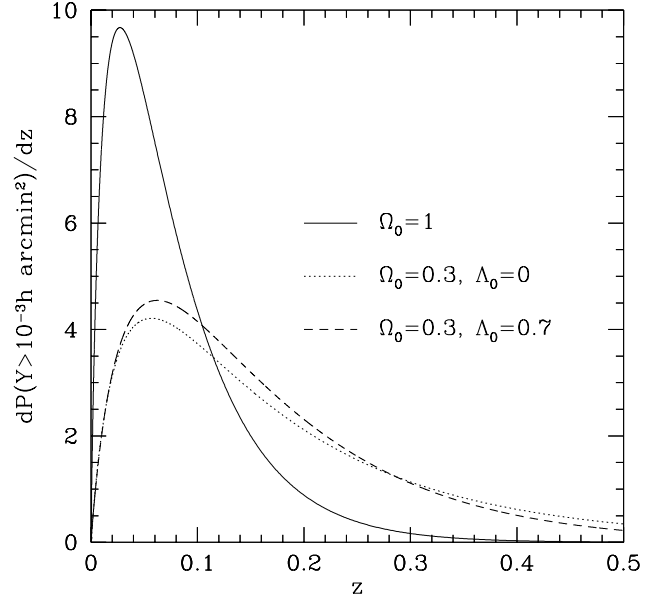}}
\caption{Redshift distribution of S-Z bright clusters
($Y>10^{-3}h$ arcmin$^2$) in different cosmological models. The ordinate
gives the probability distribution of clusters 
per unit redshift interval. The solid line corresponds to $\Omega_0=1$,
dotted line to $\Omega_0=0.3$ and dashed line to 
$\Omega_0=0.3$ and $\Lambda_0=0.7$.
}
\label{fig:pzgy}
\end{figure}

\begin{figure}
\centering
\centerline{\epsfxsize = 8.0cm \epsfbox[50 100 574 724]{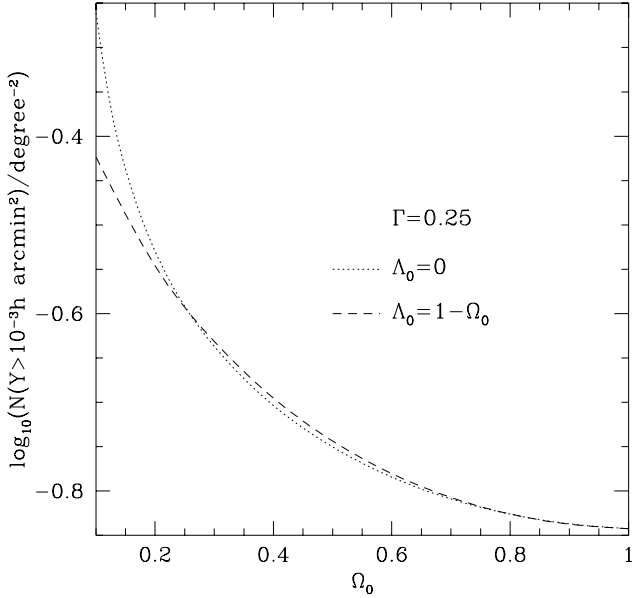}}
\caption{Predicted number counts of S-Z clusters with $Y>10^{-3}$h
arcmin$^2$ out to $z=0.33$ and $0.5$ as a function of $\Omega_0$. The
dotted line corresponds to open models ($\Lambda_0=0$) and the dashed line
to flat models ($\Lambda_0=1-\Omega_0$). The fractional difference between
the models is smaller than in the corresponding plot for the temperature
function because the counts are dominated by objects at low redshift where
the models have been normalised to match the observed temperature function.
}
\label{fig:nyzom}
\end{figure}

\begin{figure}
\centering
\centerline{\epsfxsize =8.0cm \epsfbox[50 100 574 724]{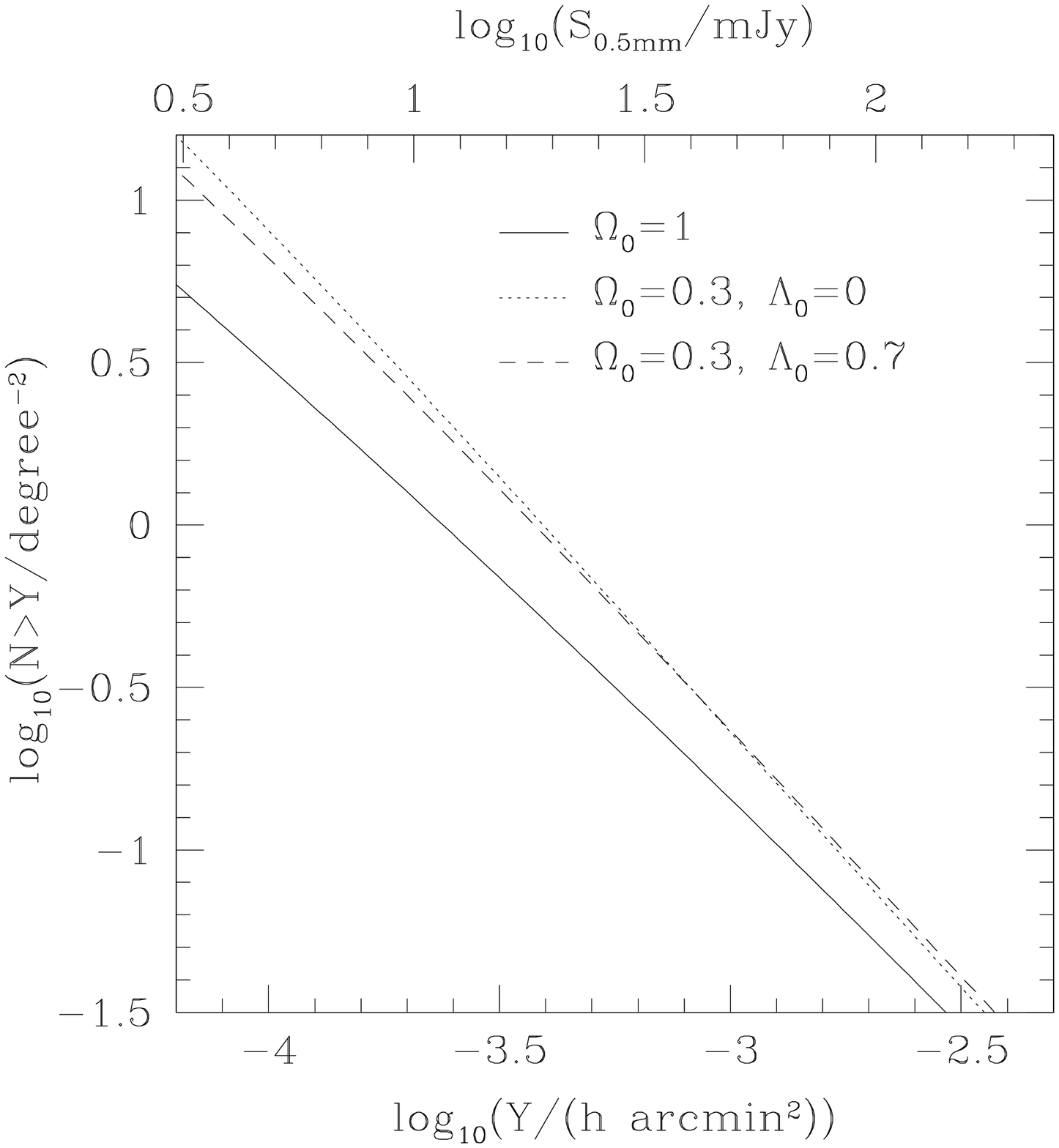}}
\caption{Sunyaev-Zel'dovich source counts as a function of $Y$ for three
different cosmologies. Solid lines correspond to $\Omega_0=1$, dotted
lines to $\Omega_0=0.3$, and dashed lines to $\Omega_0=0.3, \Lambda_0=0.7$.
The corresponding $0.5$mm fluxes in mJy are shown on the top axis.
About $60\%$ more clusters with $Y>10^{-3}h$ arcmin$^2$ are expected in the
$\Omega_0=0.3$ models than in the $\Omega_0=1$ case. 
}
\label{fig:szscount}
\end{figure}

As discussed in Section~2, we characterize the S-Z effect by the
$Y$-function given in equations~(2.5) and~(\ref{yofm}). (Here, we take
$f_{\rm ICM}=0.1$ and $\beta=1$, but our results are readily scaled to
other values using $Y \propto f_{\rm ICM} \beta^{-1}$.) The evolution of
the $Y$-function is similar to that of the temperature function, but since
$Y \propto M^{5/3}$ whilst $T \propto M^{2/3}$, the detailed behaviour is
slightly different. More importantly, the appearance of the
diameter-distance in equation~(2.5) causes $Y$ to drop off with redshift
more rapidly than $kT$.

Figure~\ref{fig:yevol} confirms the rapid decline in the $Y$-function with
redshift, resulting from the $r_{\rm d}^{-2}$ term, in all our
cosmological models (which, as before, are normalised using the local
cluster X-ray temperature function). Evolution is in the same sense as
evolution in the cluster mass and X-ray temperature functions.
Also, for $z>0$, the differences between the $Y$-functions of our various
models are of similar magnitudes. To understand why this is so we need to
consider the two factors mentioned above. In the $\Omega_0=1$ model, in 
which the fluctuation growth factor is still changing rapidly at low
redshifts, $M^{5/3}$ decreases faster with redshift than either $M$, which
is relevant to the mass function, or $M^{2/3}$, which is relevant to the
temperature function. Thus, for $\Omega_0=1$, the $Y$-function evolves
more rapidly than the mass or temperature functions. This is
partly offset, in the low-$\Omega_0$ models, by their larger
angular-diameter distance out to a particular redshift.

Because of the strong evolution apparent in Fig~\ref{fig:yevol}, the
redshift distribution of ``S-Z bright" clusters is highly peaked at
low redshift in all our models. This distribution is plotted in
Fig~\ref{fig:pzgy} for clusters with $Y>10^{-3}$ h arcmin$^2$. The
corresponding number counts as a function of $\Omega_0$ are shown in
Fig~\ref{fig:nyzom}. The counts are dominated by local clusters and,
since all the models are normalised to the local temperature function,
the counts are similar in all cases.  Nevertheless, $\sim 60\%$ more
$Y>10^{-3}$h arcmin$^2$ clusters are expected if $\Omega_0=0.3$ than
if $\Omega_0=1$.  The splitting between the various cosmologies
increases if the $Y$ threshold is reduced, as illustrated in
Fig~\ref{fig:szscount}. As the threshold is lowered, the counts probe
higher redshifts where the predictions are increasingly sensitive to
$\Omega_0$. Thus, if the threshold is taken to be $10^{-4}$h
arcmin$^2$, over twice as many clusters are expected per unit area if
$\Omega_0=0.3$ than if $\Omega_0=1$. As with the counts as a function
of mass or X-ray temperature, the counts as a function of $Y$ are
insensitive to $\Lambda_0$. The redshift distribution of a sample of
Y-selected clusters gives a more promising method of constraining
$\Omega_0$. For clusters with $Y>10^{-3}$h arcmin$^2$ 
the median redshift (see Fig.~10) 
in the two $\Omega_0=0.3$ models is approximately
$\bar z = 0.136$, which is a little over twice that of the $\Omega_0=1$ model. 
Thus optical follow up on a relatively small sample of Y-selected
clusters could easily distinguish between these two models.

\section{Discussion and conclusions}\label{sec:conc}

We have used the Press-Schecther formalism to investigate
the evolution of the population of rich galaxy clusters. Our work extends
and complements earlier work using this formalism by Evrard (1989), Henry
\& Arnaud (1991), Lilje (1992), Oukbir \& Blanchard (1992), Hanami (1993),
White \etal (1993), Barbosa \etal (1995), Hattori \&
Matsuzawa (1995), and Viana \& Liddle (1995), amongst others. We first 
verified that the Press-Schechter formula predicts the correct abundance of 
rich clusters at $z=0$ and $z=0.5$ by comparing with the results of large 
cosmological N-body simulations. The agreement is excellent, at least for 
the models we have considered: CDM-like cosmologies with spectral shape 
parameter, $\Gamma=0.25$ and with (a) $\Omega_0=1$, (b) $\Omega_0=0.3$, 
$\Lambda_0=0$ and (c) $\Omega_0=0.3$, $\Lambda_0=0.7$.

We reconsidered the problem of normalising the amplitude of mass
fluctuations on cluster scales by reference to the present day abundance of
rich clusters. From our own rederivation of the X-ray temperature
distribution of clusters, using Henry \& Arnaud's (1991) data, we found 
the following values for the {\it rms} density fluctuation in spheres
of radius $8\Mpc$, $\sigma_8$:

\begin{equation}
\sigma_8 = (0.50 \pm 0.04) \Omega_0^{-0.47+0.10\Omega_0} \ \ {\rm for} 
\ \ \Lambda_0=0 
\label{sig8nolam_1}
\end{equation}
and

\begin{equation}
\sigma_8 = (0.50 \pm 0.04) \Omega_0^{-0.53+0.13\Omega_0} \ \
{\rm for} \ \ \Omega_0+\Lambda_0=1.
\label{sig8lam_1}
\end{equation}

For $\Omega_0=1$ this estimate is independent of the shape of the power
spectrum and, for other values of $\Omega_0$, there is only a very weak
dependence. Note that $\sigma_8$ is only slightly larger if the $\Lambda$
term is non-zero. For $\Omega_0>0.2$, the difference between the flat and
open models is always less than 10\%. Note also that since the {\it rms}
fluctuation in the bright galaxy distribution is $0.96$ \cite{mad96}
in spheres of radius $8 \Mpc$, our
estimates of $\sigma_8$ imply that the biasing parameter defined as
$b=0.96/\sigma_8$ is greater than unity for open models with $\Omega_0 > 0.23$
and for flat models with $\Omega_0 > 0.27$. Models with $\Omega_0$ smaller 
than this require antibiasing, that is they require bright galaxies to be less
clustered than the mass. Our estimates of $\sigma_8$ differ slightly from
those obtained by previous authors and, in Section~4, we discussed in
detail the reasons for these differences.

The quoted uncertainties in equations~(\ref{sig8nolam_1})
and~(\ref{sig8lam_1}) represent the estimated overall 
errors in the model fits to
the observed X-ray temperature function. The main source of uncertainty
in our analysis is our modelling of the X-ray emitting intracluster
medium as a homogeneous, isothermal gas in hydrostatic equilibrium (cf
equation~2.2). There is some tentative evidence from ASCA data that the
temperature in some clusters may be declining at large radii 
\cite{asca}. On the other hand, hydrodynamic simulations show that
the isothermal assumption is a good approximation in the region where most
of the X-rays are emitted, at least in the case where cooling flows are
ignored (Evrard 1990, Tsai, Katz \& Bertschinger 1994,
Navarro \etal 1995). Our modelling also
required us to fix a value of the parameter $\beta$, the ratio of specific
galaxy kinetic energy to specific gas thermal energy, and we chose to set
$\beta=1$, consistent with the results of simulations. If
bulk motions or magnetic stresses contribute to the support of the gas, a
larger value of $\beta$ would be appropriate. In this case, our results may
be recast by scaling all temperatures inversely with $\beta$ and this would
lead to higher estimates of $\sigma_8$. Finally, the estimated gas
temperatures and our inferred values of $\sigma_8$ may also be
underestimated if small-scale inhomogeneities in the
gas distribution affect the measured X-ray spectrum.

\begin{figure}
\centering
\centerline{\epsfxsize =11.5cm \epsfbox[0 0 574 724]{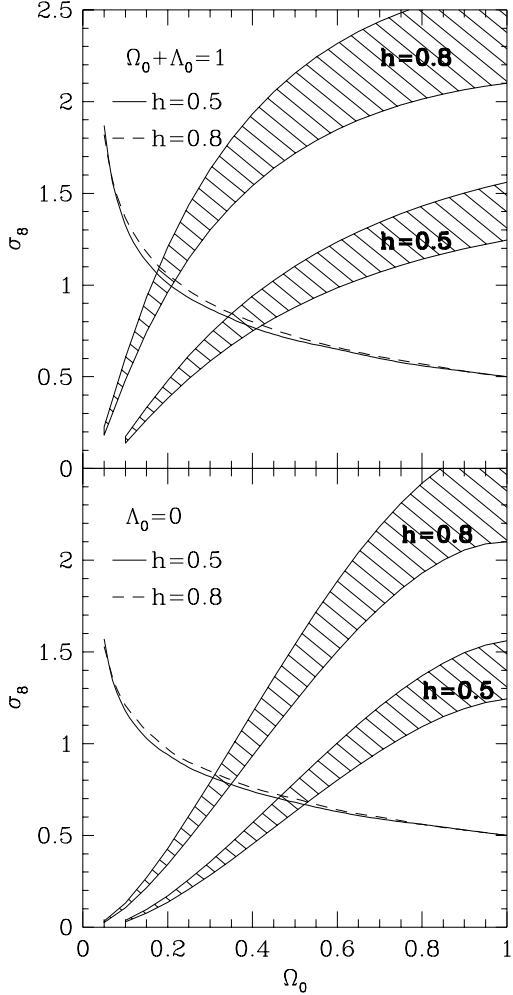}}
\caption{Comparison of $\sigma_8$ estimated from the COBE 2-year data and
from the abundance of rich galaxy clusters. The top panel corresponds
to flat models and the bottom panel to open models. The upper hatched
region shows the COBE estimates for $h=0.8$  and
the lower hatched region the estimates for $h=0.5$ 
as parameterised by Liddle \etal (1995a and~b) with
12\% errors in both cases. The dashed and solid lines show our estimates
from the cluster abundance for $h=0.8$ and $h=0.5$ respectively, obtained
through the procedure of Section~4, setting $\Gamma=\Omega_0 h~ 
{\rm exp}(-\Omega_{\rm b}-\Omega_{\rm b}/\Omega_0)$, 
where $\Omega_{\rm b}=0.013/h^2$
is the ratio of the mean baryon density to the critical density. The
statistical errors in this case are less than 5\%. 
}
\label{fig:cobe}
\end{figure}

The values of $\sigma_8$ required to match the local abundance of clusters
may be compared with measurements of the fluctuation amplitude on larger
scales, particularly with those inferred from the microwave background
anisotropies in the COBE 2-year data. By assuming a specific shape for the
fluctuation spectrum, the COBE results may be extrapolated to $8 \Mpc$ and,
in principle, comparison with equations~({\ref{sig8nolam_1})
and~({\ref{sig8lam_1}) provides a test of the assumed spectral shape. This
test, however, cannot yet be made completely rigorous because the available
anisotropy data do not distinguish between contributions from scalar modes
which determine the amplitude of the mass fluctuations, and contributions
from tensor modes which produce gravitational waves. Furthermore, the
asymptotic slope of the mass power spectrum, $n$, is poorly constrained by
the anisotropy data and this introduces further uncertainty in the analysis
of the COBE results (but not in the cluster abundance argument).  As an
illustration, we compare, in Fig~\ref{fig:cobe}, our estimates of
$\sigma_8$ with the COBE values for the simplest possible case in which $n$
takes the standard value of unity and tensor contributions are neglected.
The hashed regions in the Figure show COBE estimates for $h=0.8$ and
$h=0.5$, taken from Liddle \etal (1995a) in the open case, and
from Liddle \etal (1995b)
in the flat case. The dashed and solid lines give
our estimates of $\sigma_8$ for the same two values of $h$, derived from
the procedure described in Section~4, setting $\Gamma=\Omega_0 h 
{\rm exp}(-\Omega_{\rm b}-\Omega_{\rm b}/\Omega_0)$ \cite{sugi} with
the nucleosynthesis value of  $\Omega_{\rm b}=0.013/h^2$
\cite{copi}. 
Large values of $h$ require low values of
$\Omega_0$. Thus, if $h=0.8$, $\Omega_0$ is required to be less than $0.2$,
in the flat case, and less than $0.35$ in the open case. 
If $h>0.5$, values of $\Omega_0 \gsim 0.5$ are ruled out by this test 
whether or not $\Lambda_0=0$. 
For $\Omega_0=1$,
$h=0.27$ is required for consistency between COBE and the cluster
abundances. Thus, as noted before by
Efstathiou, Bond \& White (1992) and White \etal (1993),
the abundance of clusters in the standard CDM model ($h=0.5$) is
incompatible with the COBE fluctuations unless gravitational waves make a
significant contribution to the measured microwave background anisotropies
or there is a strong tilt in the primordial spectrum ($n\lsim 0.8$; Bond
1995).

While the present day abundance of clusters may be used to determine the
value of $\sigma_8$ with only a weak dependence on the power spectrum of
fluctuations, the evolution of the cluster abundance may
be used to set constraints on $\Omega_0$ itself. These depend sensitively
on the value of $\sigma_8$ and, to a lesser extent, on the shape of the
fluctuation spectrum. In this study, we have, for the most part, restricted
attention to models with spectral shape parameter, $\Gamma=0.25$,
consistent with observations of the large-scale distribution of galaxies
(Efstathiou \etal 1992). 
Since the $\Omega_0$ dependence of the
evolutionary rate is so strong, the main effect of changing $\Gamma$ is 
through its influence on the model normalization, $\sigma_8$. However, as
Figure~\ref{fig:cobe} shows, this is a  weak effect. 
Thus, in practice, this test of $\Omega_0$ is insensitive to small
departures from our assumed value of $\Gamma$. 
Furthermore, for interesting
values of $\Omega_0$, the diagnostics we have considered depend only very
weakly on $\Lambda_0$. Thus, statistical studies of clusters at
intermediate redshifts offer an excellent prospect for determining
$\Omega_0$ without the complications arising from the uncertain shape of
the galaxy power spectrum and the poorly understood 
relation between the distributions of galaxies and mass.

Different observables may be used to characterize the evolution of the
cluster population. Here we have considered three properties for which
observational data are likely to be obtained in the near future: the
distribution of cluster masses, X-ray temperatures and S-Z decrements.  The
first may be derived from weak gravitational lensing studies; the second
from existing and forthcoming X-ray surveys; and the last from ground-based
observations and proposed space missions such as COBRA/SAMBA. Although the
masses are perhaps the most difficult to measure, they provide a
particularly robust test since, apart from the Press-Schechter ansatz, the
only model assumption is the form of the density distribution of dark
matter halos. 
In rough agreement with N-body simulations, we have taken 
this to be a singular isothermal sphere. In all cases, we have presented
expected distributions at $z=0.33$ and $z=0.5$ and the predicted number
counts of the largest clusters, both in space and in projection on the sky,
as a function of redshift.

In agreement with previous analyses, we find that even at redshifts as low
as 0.33, the comoving abundance of high mass or high X-ray temperature
clusters is radically different in universes with $\Omega_0=1$ and 
$\Omega_0=0.3$. For example, at this redshift we expect 20 times as many
clusters with $M>3.5 \times 10^{14}\Msol$ and 5 times as many clusters with
$kT>5$ keV if $\Omega_0=0.3$ than if $\Omega_0=1$. The corresponding
factors at $z=0.5$ are $\sim$ 65 and 15. 
When integrating these cluster number densities with respect to redshift
the effects of the different volume elements are included.
Thus, the expected numbers of clusters per
square degree on the sky out to redshift 0.5 differ by a factor of $\sim$ 10
for both $M>3.5 \times 10^{14}\Msol$ and $kT>5$ keV in these two
cosmologies.

Whereas the mass and temperature functions only evolve dramatically if
$\Omega_0=1$, the distribution of S-Z decrements (as measured by the
$Y$-function of equation~\ref{yofm}) declines rapidly with redshift in
all cosmologies. This is mainly because, for the range of redshifts
that we have considered, an object far away subtends a smaller angle
on the sky than the same object placed nearby. Even so, at $z=0.33$,
the comoving number density of clusters with $Y=10^{-3}$h arcmin$^2$
is 10 times higher if $\Omega_0=0.3$ than if $\Omega_0=1$. However,
the overall abundances at this redshift are all very low because of
the rapid decrease in the angular size of the clusters with
redshift. The projected counts are dominated by low redshift clusters
and, as a result, the expected excess of clusters with $Y>10^{-3}$h
arcmin$^2$ is only $\sim 60\%$ if $\Omega_0=0.3$. This excess becomes
larger as the $Y$ threshold is lowered and higher redshift clusters
are included.

To obtain an estimate of $\Omega_0$ using any of the tests proposed in
this paper requires complete samples of intermediate redshift clusters
selected according to the statistic under consideration. Samples of
S-Z clusters are likely to be $Y$-limited in any case.  For the other
diagnostics, selecting by X-ray luminosity is probably an efficient
method. Since X-ray luminosity correlates reasonably well with X-ray
temperature (at least locally) and, apparently also with lensing mass
\cite{smail}, the sample completeness with respect to the
relevant statistic may be determined {\it a posteriori} from these
data. Incomplete datasets, on the other hand, may be used to set {\it
upper limits} on $\Omega_0$ because the abundance of clusters at
intermediate redshift declines monotonically with increasing
$\Omega_0$.

There are already several X-ray bright clusters known at high
redshift. For example, Luppino \& Gioia \shortcite{lup} report the
detection of 6 such clusters with $z=0.5$ in the EMSS survey
\cite{emss}. Measurements of their X-ray temperatures would be
extremely valuable. They might confirm the indication from velocity
dispersion measurements that these clusters are hot, suggesting that
$\Omega_0$ is low, or they might show that these velocity dispersions
are overestimated as a result of contamination by projection effects.
The lowest velocity dispersion reported for these clusters corresponds
to a temperature of 6.6 keV.  As an illustration, if we assume that
the effective area surveyed to find these six clusters was $200$
square degrees, then we would expect to find $< 0.04$ clusters above
this temperature if $\Omega_0=1$; $\sim 6$ if $\Omega_0=0.3$ and
$\Lambda_0=0.7$; and $\sim ~30$ if $\Omega_0=0.3$ and
$\Lambda_0=0$. However, if the coolest of these clusters actually has
a temperature of only 3 keV, then we would have expected $\sim 30$
examples even if $\Omega_0=1$.

In summary, a statistical sample of X-ray clusters at intermediate redshifts
with measured temperatures or S-Z decrements could place a strong
constraint on the density parameter $\Omega_0$.

\section*{ACKNOWLEDGMENTS}

We thank Hugh Couchman for providing a copy of his excellent AP$^3$M
N-body code and for giving valuable advice and support. We thank David
Weinberg for generously allowing us first use of this set of N-body
simulations and Pat Henry for his helpful advice and comments on the
manuscript. VRE acknowledges support of a PPARC studentship and SMC of a
PPARC Advanced Fellowship. This work was also supported by a PPARC rolling
grant.

\bigskip

\appendix
\section[]{The spherical collapse model for $\Omega+\Lambda=1$}

It can be shown \cite{peeb2} that the variation
of the radius, $r$, of a uniform spherical overdensity 
with scale factor $a$ is described, for a flat cosmology with a cosmological
constant, by
\begin{equation}
 \left(\frac{{\rm d}r}{{\rm d}a}\right)^2 = \frac{r^{-1}+\omega r^2-\kappa}
 {a^{-1}+\omega a^2},
\label{drda}
\end{equation}
where $a=(1+z)^{-1}$, $\kappa$ is a
constant which for overdensities takes positive values
and
\begin{equation}
\omega=(\Omega_0^{-1}-1)  .
\label{omega}
\end{equation}
Therefore a perturbed region will turnaround when
\begin{equation}
 \omega r^3 - \kappa r + 1 = 0.
\label{ta}
\end{equation}
Solving this cubic and requiring that a physically sensible root exists 
we find the smallest perturbation which will collapse has
\begin{equation}
 \kappa_{\rm min} = \frac{3\omega^{\frac{1}{3}}}{2^\frac{2}{3}} .
\label{kmin}
\end{equation}
The corresponding turnaround radius is
\begin{equation}
 r_{\rm ta,max} = (2\omega)^{-\frac{1}{3}}.
\label{rtamax}
\end{equation}
Perturbations with larger values of $\kappa$ turnaround and collapse
earlier. Solving the cubic (\ref{ta}) we find the turnaround radius
$r_{\rm ta}$ as a function of the density parameter $\omega$ and 
perturbation amplitude $\kappa$. The solution can be expressed 
as
\begin{equation}
 r_{\rm ta} = -2s^{\frac{1}{3}}\cos\left(\frac{\theta}{3}-\frac{2\pi}{3}\right),
\label{rta}
\end{equation}
where
\begin{equation}
s=\left(\frac{3}{2}\right)^3\left(\frac{\kappa}{\kappa_{\rm min}^3}\right)
^{\frac{3}{2}}
\label{s}
\end{equation}
and $\theta$ satisfies
\begin{equation}
\cos \theta =
\left(\frac{\kappa_{\rm min}}{\kappa}\right)^{\frac{3}{2}}
\hspace{20pt} \left( 0 < \theta < \frac{\pi}{2} \right).
\label{costh}
\end{equation}

In order to calculate the redshifts corresponding to turnaround
and collapse we separate the variables in (\ref{drda}) 
and integrate,
\begin{equation}
\int^{r_{\rm ta}}_0 \frac{r^{\frac{1}{2}}}{(\omega r^3 - \kappa r + 1)^{\frac{1}{2}}}{\rm d}r = \int^{a_{\rm ta}}_0 \frac{a^{\frac{1}{2}}}
{(\omega a^3 + 1)^{\frac{1}{2}}}{\rm d}a.
\end{equation}
Defining
\begin{equation}
I(\omega,\kappa) =\int^{r_{\rm ta}}_0 
\frac{r^{\frac{1}{2}}}{(\omega r^3 - \kappa r + 1)^{\frac{1}{2}}}{\rm d}r ,
\label{defi}
\end{equation}
which must be evaluated numerically, the expansion factor 
at turnaround, $a_{\rm ta}$, and at collapse, $a_{\rm c}$, satisfy
the implicit equations,
\begin{equation}
{\rm I(\omega,\kappa}) = \int^{a_{\rm ta}}_0 \frac{a^{\frac{1}{2}}}
{(\omega a^3 + 1)^{\frac{1}{2}}}{\rm d}a
\end{equation}
and
\begin{equation}
2\, {\rm I(\omega,\kappa}) = \int^{a_{\rm c}}_0 \frac{a^{\frac{1}{2}}}
{(\omega a^3 + 1)^{\frac{1}{2}}}{\rm d}a ,
\end{equation}
where the collapse time is defined in the usual way to be that at which 
(\ref{drda}) predicts collapse to a point singularity.
Using the substitution $u = \sqrt{\omega a^3}$, 
these equations can be integrated analytically to give
\begin{equation}
a_{\rm ta} = \left[\frac{\exp({3\sqrt{\omega}{\rm I(\omega,\kappa)}}) - 1}
{2\sqrt{\omega}}\right]
^{\frac{2}{3}} \exp(-\sqrt{\omega}  I(\omega,\kappa))
\label{ata}
\end{equation}
and
\begin{equation}
a_{\rm c} = \left[\frac{\exp({6\sqrt{\omega}~{\rm I(\omega,\kappa)}}) - 1}
{2\sqrt{\omega}}\right]
^{\frac{2}{3}} \exp(-2 \sqrt{\omega}  I(\omega,\kappa))
\label{ac}
\end{equation}
Hence we have $a_{\rm c}$, or equivalently $z_{\rm c}$, 
the redshift of collapse of a spherical
overdensity, as a function of $\omega$ and $\kappa$.

To calculate $\Delta_{\rm c}$, the density of virialised object in terms of
critical density, we make use of the expression from \cite{lahav}
which gives the ratio between the radius of the virialised sphere and its
turnaround radius. Given the expression above for the 
turnaround radius and the expansion factor at collapse, $\Delta_{\rm c}$
is given by
\begin{equation}
\Delta_{\rm c}=\Omega(a_{\rm c})\left(\frac{a_{\rm c}}{r_{\rm c}}\right)^3.
\label{delc}
\end{equation}

All that remains is to relate $\kappa$ 
to the quantity more often used to describe
a perturbation, namely $\delta_0$,
the linear theory overdensity extrapolated to the present.
This in turn is related to the linear theory overdensity
at collapse, $\delta_{\rm c}$, by
\begin{equation}
\delta_{\rm c} = \delta_0 D(a_{\rm c})
\label{dc}
\end{equation}
where $D(a)$ represents the linear theory growth factor
normalised to unity at the present, $D(a_0)=1$. 
For the 
$\Omega_0+\Lambda_0=1$, $\Omega_0<1$ case
\begin{equation}
D(a)=\frac{A(a (2\omega)^{1/3})}{A(a_0(2\omega)^{1/3})} ,
\end{equation}
where
\begin{equation}
A(x) = \frac{(x^3 + 2)^{\frac{1}{2}}}{x^{\frac{3}{2}}}\int^{x}
\left(\frac{u}{u^3 + 2}\right)^{\frac{3}{2}}{\rm d}u
\end{equation}
\cite{peeb}.
In the limit where $a \ll 1$ we find
\begin{equation}
\delta = \frac{\delta_0 a_0(2\omega)^{1/3}}
{5 A(a_0(2\omega)^{1/3})}a + \rm{O}(a^2),
\end{equation}
which can be compared to the corresponding expression 
\begin{equation}
\delta = \frac{3\kappa}{5}a + \rm{O}(a^2),
\end{equation}
derived from (\ref{drda}).
Identifying terms we find
\begin{equation}
\kappa = \frac{a_0 (2\omega)^{1/3}}{3A(a_0 (2\omega)^{1/3})} \,\delta_0.
\label{kappa}
\end{equation}
Now given the density parameter $\Omega_0$ and
the amplitude of a density perturbation, $\delta_0$ we can compute
$\omega$ using (\ref{omega}) and $\kappa$ using (\ref{kappa}).
Then the collapse redshift can be obtained from 
(\ref{rta}),(\ref{defi}) and (\ref{ac}). The corresponding values
of $\delta_{\rm c}$ and $\Delta_{\rm c}$ are then given by
(\ref{dc}) and (\ref{delc}).
The dashed lines in Figure~\ref{fig:scoll}
show $\delta_{\rm c}$ and $\Delta_{\rm c}$ as
functions of $\Omega$ at the time of collapse.

\end{document}